\documentstyle[12pt]{article}
\begin{document}
                                        \begin{titlepage}
\begin{flushright}
DESY 94--179\\
PSU/TH/150\\
hep-ph/9410271\\
October 1994\\
\end{flushright}
\vskip0.8cm
\begin{center}
{\LARGE
Instanton-induced production of  QCD jets
  }

           \vskip1cm
{\Large I.I. Balitsky}~$^\ast$ \\
\vskip0.2cm
        Physics Department, Penn State University,\\
       104 Davey Lab.,University Park, PA 16802, USA\\
\vskip0.5cm
and\\
\vskip0.5cm
 {\Large V.M.~Braun} $\footnote { On leave of absence from
St.Petersburg Nuclear
Physics Institute, 188350 Gatchina, Russia}$ \\
\vskip0.2cm
       DESY, Notkestra\ss e 85, D--22603 Hamburg, Germany

\vskip1cm
{\Large Abstract:\\}
\parbox[t]{\textwidth}{
We consider the instanton contributions for the production of a
gluon jet with large transverse momentum in QCD. We find
 that Mueller's corrections corresponding to the
rescattering of hard quanta are likely to remove contributions of
large instantons, making this cross section well defined.
This observation generalizes the previous discussion of instanton
 effects in the deep inelastic scattering and
suggests that all hard processes in the QCD
receive hard non-perturbative corrections from
instantons of small size, of order $\rho\sim 1/(Q\alpha_s)$.
}
\\ \vspace{1.0cm}
{\em submitted to Physics Letters B }
\end{center}
                                                \end{titlepage}

\newpage
{\bf\large 1.}\hspace{0.5cm}
Gauge theories such as the QCD are known to possess a rich structure of the
scattering matrix which reflects the
disappearance of quark and gluon singularities and appearance of numerous
poles and cuts corresponding to hadrons.  It is a common
belief that this complex structure reveals itself in full
 at small energies,
while at large energies (in hard processes) all the influence of the rich
texture of QCD reduces to a few phenomenological characteristics such
as parton densities. Given these densities all properties of a given hard
process can be calculated within perturbative QCD.
Even at small distances, however, there
is a remnant of the nontrivial QCD vacuum structure which can lead to
non-perturbative contributions to hard amplitudes ---
 the small-size instantons.

In recent years there is a revival of interest to instanton
effects in gauge theories, inspired by the conjection \cite{ring90}
that in high-energy collisions such effects are enhanced enough to
produce an observable violation of the baryon number.
Although present theoretical arguments rather seem to disfavour such
a strong enhancement \cite{matt92}, so that
the instanton effects in the electroweak theory presumably
remain far below the level at which
they might become observable at future colliders, these studies
have trigged an increasing  interest to semiclassical
effects in gauge theories at high energies in a more general context.

In \cite{bal93a,bal93b} we have suggested to study the
instanton contributions in high-energy collisions in the QCD.
In this case the coupling is not so small as in the electroweak
theory, and the instanton contributions might be observable
even if they remain exponentially supressed. The Ringwald's
phenomenon --- enhancement of the cross section through the
dominance of final states with many gluons ---
allows one to hope that these cross sections may reach observable
values, and similtaneously provides a good trigger for their
observation, since a fireball of $\sim 2\pi/\alpha_s$ gluons
is likely to produce an event with a very high density of particles
in the final state.

The major difficulty for the identification of instanton
effects in QCD is that in the generic situation they are not
infra-red (IR) stable, typically involving
a IR-divergent integral over the instanton sizes.
There are speculations that in the true QCD vacuum this
integration is effectively cut off at sizes of order $\rho\sim
1/600$ MeV, but this assertion is difficult to justify
theoretically.
To be on  a safer side, one should take special care
to select contributions of small instantons, which implies
going over to a certain hard process.
This notion is applied to
reactions in which hard scale is related either
to a large virtuality
of the external particle (photon or W,Z boson),  or
to a large momentum transfer.
 In perturbation theory this
distinction is subtle: a large momentum transfer necessarily
involves an exchange by a highly-virtual gluon (quark).
Thus, perturbative description of these processes
is similar, in both cases
dynamics of small distances can be factorized from
the large distance effects. It is this way that
most of the QCD predictions arise.

This distinction is crucial, however, for the discussion of
instanton effects.
In the deep inelastic lepton-hadron
scattering the
hard scale is brought in by the virtuality $Q^2$ of the photon.
In this case \cite{bal93a,bal93b} the
contribution of small instantons is distinguished
by a non-trivial power dependence on $Q$, corresponding
to a fractional twist, and can be
disentangled from IR divergent contributions
of large instantons, included in parton distributions.
This is largely due to the fact that the instanton field at large
virtualities contains a gauge-independent piece proportional to
$\exp(-\rho|Q|)$ \cite{KR}.  Thus,
coefficient functions in front of parton distributions receive
well-defined non-perturbative corrections
\begin{equation}\label{CF}
C(x,Q^2)\sim
\exp\Bigg\{-\frac{4\pi}{\alpha_s(\rho)}
\left[1-\frac{3}{2}\left(\frac{1-x}{1+x}\right)^2+\ldots\right]
\Bigg\}\,,
\end{equation}
coming from instantons with the size of order
 $           \rho \sim \pi/\alpha_s\cdot{1}/{Q}$.
These corrections have been calculated in \cite{bal93b} and
turn out to be of order $10^{-2}-10^{-5}$
in the region of sufficiently large $Q^2>100$ GeV$^2$
and Bjorken variable $x>0.3$, where the derivation of (\ref{CF})
is justified.

The situation proves to be essentially different in large
momentum transfer reactions, from which we consider
production of a gluon jet with large $q_\perp$    as a
representative example. On physical grounds it
is obvious that this cross section cannot be affected by large
instantons.
However,
a semiclassical calculation
fails in this case: to the accuracy to which (\ref{CF}) has been derived
in \cite{bal93b},
the instanton contribution is given by a power-like divergent
integral, and contributions of small instantons $\rho\sim 1/q_\perp$
do not produce any non-trivial dependence on $q_\perp$.
Indeed, to the semiclassical accuracy the effect of small instantons
is to introduce
new {\em point-like } multi-particle vertices, which do not
involve
any momentum transfer dependence. Thus, instanton-induced amplitudes
do not decrease at large momentum transfers.

Conceptually, it is easy to realize what is missing: to obtain a
sensible result one
must take into account an (exponentially small) overlap
between the initial state, which involves a few hard quanta, with the
semiclassical final state \cite{banks}.
This necessarily involves taking into account quantum corrections
to semiclassical amplitudes in the instanton background, the study
of which has been pioneered by Mueller \cite{MU91}, see \cite{matt92}
for a review and further references.
We demonstrate that the ``Mueller's corrections" indeed
remove contributions of large instantons to the
jet production with large $q_\perp$, making the
non-perturbative contribution to this cross section well defined.

{\bf\large 2.}\hspace{0.5cm}
Mueller finds \cite{MU91} that to the one-loop accuracy the
asymptotics of the gluon propagator in the instanton background
takes the factorized form
\begin{equation}\label{propagator}
   G(p,q) \simeq A_I(p) A_I(q) \frac{\alpha_s}{8\pi}(pq)\ln(pq)
\,,
\end{equation}
assuming $pq\gg 1/\rho^2$, $p^2=q^2=0$. From this expression,
 it is possible to derive that the corresponding quantum
correction to the instanton-induced amplitude acquires the
multiplicative factor
\begin{equation}\label{Mgeneral}
   \exp\left\{-\frac{\alpha_s}{8\pi}M(p_i)\right\}\,,
\end{equation}
where
\begin{equation}\label{Msum}
 M(p_i) = \sum_{i<j} (p_i\cdot p_j) \ln (p_i\cdot p_j)\,.
\end{equation}
The summation goes over all the ingoing and outgoing particles.
To $O(\alpha_s)$ accuracy this formula follows directly
from (\ref{propagator}), provided $(p_i\cdot p_j)\gg 1/\rho^2$,
while the
exponentiation of this result is a plausible
conjecture beyond,
which has created a lot of discussions, see \cite{matt92}.
Our strategy in this work
 is to take the quantum correction in (\ref{Mgeneral})
for granted, and try to understand its qualitative
effect on the jet production.\footnote{In what follows
we neglect imaginary parts of the logarithms in (\ref{Msum}) which
produce a phase factor and to our
accuracy cancel in the answer for the cross section.}
We note in passing that in the deep inelastic scattering
the quantum correction (\ref{Mgeneral}) is
necessary to cancel the ambiguity in the $\bar I I$ interaction.
In this case the instanton size determined from the saddle-point
equations is of order
$\rho\sim(4\pi)/\alpha_s\cdot 1/Q\cdot 1/\xi^2$ where
$\xi\simeq(R/\rho)^2$ \cite{bal93a}, and the $\bar I I$ separation
$R$ is fixed by the Bjorken $x$ alias by the initial energy.
At large $\xi\gg 1$ one has, generally, $M\sim Q^2$ and thus the
expression under the exponent in (\ref{Mgeneral}) is of order
$\pi/\alpha_s \cdot 1/\xi^4$, same as ambiguities in
$U^{\bar I I}_{\rm int}$.

{\bf\large 3.}\hspace{0.5cm}
 The Mueller's factor $M$ (\ref{Msum}) involves momenta
of hard particles (a few), which we denote by $p_i$, and soft
particles ($n\sim (4\pi)/\alpha_sU^{\bar I I}_{\rm int}$),
denoted by $k_i$.
Hard particles are the two colliding partons (gluons) and the
final state gluons with large momentum, which are resolved as
jets. We consider the inclusive cross section, summing over
all soft particles and integrating over their
phase space. As it is well known, an explicit integration over the
final phase space can be rewritten in terms
of the instanton-antiinstanton interaction, which is
exemplified by writing
\begin{eqnarray}\label{qspace1}
e^{-(4\pi)/\alpha_s(1-6/\xi^2)} &=&
e^{-2 S_0}\exp\Big\{\frac{24\pi}{\alpha_s}\frac{\rho_1^2\rho_2^2}{R^4}\Big\}
\nonumber\\
&=&e^{-2S_0}\exp\Big\{-\frac{16\pi^3}{\alpha_s}\rho_1^2\rho_2^2
\int\frac{dk}{(2\pi)^4}\, e^{-ikR}\frac{(k\cdot E)^2}{E^2 k^2}\Big\}\,,
\end{eqnarray}
where $S_0=2\pi/\alpha_s$ is the instanton action,
$E$ is the total momentum of soft particles, and the maximal attractive
$\bar I I$ orientation is assumed.
 For further use we have introduced the notation
$
\xi ={R^2}/({\rho_1\rho_2})
$.
 Here $\rho_1$ and $\rho_2$ are the
sizes of the instanton and antiinstanton, and $R$ is their separation which
we take to be  parallel to $E$.
 The restriction to the special cases
 of maximal attractive $I\bar{I}$ orientation
in color space and $R\parallel E$ orientation in normal space
are justified $a~posteriori$ when it turns out that the final
integral over the collective coordinates of instantons is determined
by the saddle point where these two properties are valid
(see \cite{bal93a} for details).

Eq.(\ref{qspace1}) corresponds to the approximation when all interactions
between the soft particles in the final state are neglected. Taking them
into account corresponds to the substitution of the first two terms
in the expansion $S(\xi) = 1-6/\xi^2+\ldots$ by the full expression
$S(\xi)$ for  the QCD action on the $\bar I I$ valley configuration \cite{by}.

The Mueller's correction generally does not allow to eliminate the
explicit dependence on momenta of soft particles,
 because of
the terms corresponding to ``soft-hard'' interactions, which involve
particles with both the hard (initial and final) and soft momenta.
We are going to demonstrate, however, that at large values
of the conformal parameter $\xi$
the soft momenta appearing in $M$ can be
substituted by
\begin{equation}\label{sub}
     k_i \rightarrow \frac{E=\sum k_i}{n}\,.
\end{equation}
where $n$ is the number of soft particles which eventually disappears
in the final expressions. The total energy of soft particles $E$ is
determined by the energy conservation.\footnote{
It is convenient to use the
same notation $E$ for the four-vector of the sum of the soft particle
momenta and for their energy in the c.m. frame, which, hopefully, cannot
produce a confusion.}
Thus, to this accuracy $M$
is expressed entirely in
terms of hard momenta.

The idea of the derivation is to use a freedom
 to introduce an arbitrary common
factor under the logarithms in (\ref{Msum}), see \cite{MU91},
 to choose this factor
in such a way that
the terms quadratic in soft momenta become small and can be neglected.
To this end, we write down
\begin{equation}\label{M1}
 M(p_i) = \sum_{i<j} (p_i\cdot p_j) \ln (\rho^2 p_i\cdot p_j)
        -\sum_{i,j} (p_i\cdot k_j) \ln (\rho^2 p_i\cdot k_j)
        +\sum_{i<j} (k_i\cdot k_j) \ln (\rho^2 k_i\cdot k_j)\,,
\end{equation}
separating contributions of hard and soft momenta explicitly and
introducing the scale $\rho$ under the logarithms. To our accuracy it is
enough to take
$\rho_1\sim\rho_2\sim\rho$. The three terms in (\ref{M1}) correspond to
``hard-hard'', ``soft-hard'' and ``soft-soft'' corrections, respectively.
The derivation of (\ref{Msum}) is justified for the first two terms,
involving large relative momenta, while the third piece should be absorbed
in the full ``soft-soft'' correction. The crucial observation, which
will be justified {\em a posteriori}, is that in the form in (\ref{M1})
(that is after the proper scale is included under the logarithms) the
third term is small, of order $\sim 1/\xi^3$ compared to the first
two terms, and can be neglected.

Restricting to the  ``soft-hard'' terms only, the instanton-induced
cross section can be written, apart from the $\rho$-integration, as
 \begin{eqnarray}
\sigma_I &\sim & \int dR\,e^{iER}\exp\Bigg\{
-\frac{16\pi^3}{\alpha_s}\rho_1^2\rho_2^2\int\frac{dk}{(2\pi)^4}
\frac{(k\cdot E)^2}{E^2 k^2}
\nonumber\\
&&{}\times\exp\Big[-ikR-\frac{\alpha_s}{8\pi}(\rho_1^2+\rho_2^2)
        \sum_{i,j} (p_i\cdot k_j) \ln (\rho^2 p_i\cdot k_j)\Big]\Bigg\}\,.
\end{eqnarray}
As we shall soon demonstrate, large values of $\xi$
correspond to a small fraction
of the total energy transferred to the instanton, of order
${E^2/ s}\sim \xi^{-3}$, (see eq. (\ref{saddle}) below).
Since by assumption $R\parallel E$,
we find that
$$\frac{(p_i\cdot p_j)R^2}{(p_i\cdot R)(p_j\cdot R)}\sim \xi^{-3}\,,$$
because typically $(p_i\cdot E)\sim(p_i\cdot p_j)\sim s$
whereas $E^2\sim {s/ \xi^3}$. Using this small parameter,
the internal integral over $k$ can be taken and gives:
\begin{eqnarray}
\lefteqn{\int\frac{dk}{(2\pi)^4}\frac{(k\cdot E)^2}{E^2 k^2}
\exp\Big[-ikR-\frac{\alpha_s}{8\pi}(\rho_1^2+\rho_2^2)
        \sum_{i,j} (p_i\cdot k_j) \ln (\rho^2 p_i\cdot k_j)\Big]=}
\nonumber\\
&=&{}-\frac{3}{4\pi^2}\Big[R-i\frac{\alpha_s}{8\pi}
\rho^2\sum_i p_i\ln(-2p_i\cdot R)\Big]^{-4}\hspace{4.5cm}{}
\end{eqnarray}
Making the shift of the integration variable in the remaining integral
over the $\bar I I$ separation
$$R\rightarrow R' = R-i\frac{\alpha_s}{8\pi}
\rho^2\sum_j p_j\ln(-2p_j\cdot R)$$
we obtain
\begin{equation}\label{qspace2}
\sigma_I \sim \int dR\,\exp\Big[iER -\frac{\alpha_s}{8\pi}(\rho_1^2+\rho_2^2)
\sum_j (p_j\cdot E)\ln(p_j\cdot E)-\frac{4\pi}{\alpha_s}S(\xi)\Big]\,.
\end{equation}
It is easy to see that the remnant of the ``soft-hard'' interaction is
exactly of the form corresponding to the substitution in (\ref{sub}).
Indeed, inserting (\ref{sub}) in (\ref{M1}) and assuming $n\gg 1$ one gets
\begin{eqnarray}\label{M2}
\lefteqn{ M(p_i) =}
\nonumber\\
&= &\sum_{i<j} (p_i\cdot p_j) \ln (\rho^2 p_i\cdot p_j)
        -\sum_{i} (p_i\cdot E) \ln (\rho^2 p_i\cdot E/n)
        + \frac{n(n-1)}{2}(E/n)^2 \ln (\rho^2 (E/n)^2)
\nonumber\\
&=&\sum_{i<j} (p_i\cdot p_j) \ln (\rho^2 p_i\cdot p_j)
        -\sum_{i} (p_i\cdot E) \ln (\rho^2 p_i\cdot E)
        + E^2 \ln (\rho^2 E^2)\,,
\end{eqnarray}
where we have used the momentum conservation implying
$\sum_{i} p_i =E$.
The second term in (\ref{M2}) reproduces (\ref{qspace2}),
and the last term can now be neglected, since
it is of the order $\sim 1/\xi^3$, see the second of eqs.(\ref{saddle}) below.

{\bf\large 3.}\hspace{0.5cm}
In this approximation, it is straighforward to
 calculate the  Mueller's factor for particular processes.
For the back-to-back production of a
pair of gluon jets $ g+g \rightarrow g+g+X$  we find
\begin{equation}
   M= 2 \ln 2 (p-q)^2 + 4 pq \ln 2 \,G(\theta)\,,
\end{equation}
\begin{equation}
\ln 2\,
G(\theta) = -\sin^2\frac{\theta}{2} \ln\sin^2\frac{\theta}{2}
    - \cos^2\frac{\theta}{2} \ln\cos^2\frac{\theta}{2}\,,
\end{equation}
where $p,q$ are the momenta of ingoing and outgoing gluons in c.m.
frame,
respectively, and $\theta$ is the angle between them.
The instanton-induced cross section to the exponential accuracy
reads
\begin{equation} \label{sigma}
\sigma_I\sim \int dR\,d\rho\, \exp\Bigg\{ER-\frac{4\pi}{\alpha_s} S(\xi)
 -\frac{\alpha_s}{4\pi}M(p,q)\rho^2\Bigg\}\,.
\end{equation}
Here $E=2(p-q)$ is the energy transferred to the instanton
(and released in soft particle emission).

The integral is taken by the saddle-point method.
The saddle-point values for $\rho$ and $R$ are determined from the
equations
\begin{eqnarray}
 E\rho &=& \frac{8\pi}{\alpha_s}\sqrt{\xi-2}S'(\xi)\,,
\nonumber\\
 M \rho^2\left[\frac{\alpha_s}{\pi}+\frac{b}{4}
\left(\frac{\alpha_s}{\pi}\right)^2\right] &=&
\frac{16\pi}{\alpha_s}(\xi-2)S'(\xi)+4 b S(\xi)\,,
\end{eqnarray}
where $S'(\xi) = (d/d\xi)S(\xi)$ and $b=11-(2/3) n_f$.
Neglecting for
simplicity the running of the QCD coupling and taking into account
the dipole interaction term in the expansion of the action
$S(\xi) = 1-6/\xi^2+\ldots$, one gets
\begin{equation}\label{saddle}
 \frac{\alpha_s}{\pi} \rho = \frac{2E}{M}\sqrt{\xi}\,,~~
 \frac{E^2}{M} = \frac{48}{\xi^3}\,,
\end{equation}
justifying the assumptions which we have made to calculate the
``soft-hard'' corrections.

Now comes the central point. The function $G(\theta)$ varies
between 0 and 1, with a minimum value $G=0$ at $\theta=0$ and
$\theta=\pi$, and a maximum $G=1$ at $\theta=\pi/2$.
Consider first the collinear jet production, $\theta=0,\pi$.
Hence $M=2 \ln 2 (p-q)^2 = E^2/2\cdot\ln 2$ is of order of the energy
transferred to the instanton. From the second of the saddle-point
equations in (\ref{saddle}) one finds then $\xi^3 = 24 \ln 2$,
independent on the external momenta. Thus, in this
case the cross section is defined by the region of $R\sim \rho$ where
instantons interact strongly and the calculation is not justified
(parametrically). On the other hand, consider jets with
large transverse momentum, $\theta = \pi/2$. Then $M=2\ln 2 (p^2+q^2)
\simeq s \ln 2$, where $s=4 p^2$ is the total energy, and
substituting this to the saddle-point equation we find
$   \xi^3 = 48 \ln 2 \cdot s/E^2$.
Keeping $E^2\ll s$ (which means that momenta of gluon jets are
close to momenta of coliding gluons) we get $R\gg\rho$ and
the calculation is under control.
Note that we get $\alpha_s/4\pi\cdot \rho^2 M \sim \pi/\alpha_s
\cdot 1/\xi^2$, which indicates that the Mueller's correction
now contributes on the leading $1/\xi^2$ level, same as the
dipole interaction.

The same effect is observed in the instanton-induced production
of monojets $g+g\rightarrow g+X$, in which case for $\theta=\pi/2$
in the c.m. frame we obtain
\begin{eqnarray}
M &=& s \ln 2 -(s/2) \ln(1+E^2/s)+
(E^2/2) \ln 2 \nonumber\\ &&{}
- (E^2/2)\ln(1+s/E^2)\,.
\end{eqnarray}
The cross section is given again by the integral in (\ref{sigma}),
and the saddle-point values in the limit $E^2/s \ll 1$ are
\begin{equation}
   \xi^3 = 48 \ln 2\, s/E^2 \,,~~
   \sqrt{s}\,\rho  =
    \frac{4\pi}{\alpha_s}\cdot \frac{\sqrt{3\ln 2}}{\xi}\,.
\end{equation}
 Thus, at least in this academic limit, the calculation is
under control. This example can be interesting phenomenologically,
since it has a clear signature and smaller perturbative background.
Typical numbers are as follows: in gluon-gluon collisions with
$\sqrt{s}=400$ GeV, one could look for production of a monojet
with $q_\perp>180$ GeV, balanced by $n_g\sim 10-15$ gluons and
$2n_f$ quarks with transverse momenta of order $\rho^{-1}\sim
10$ GeV each. The cross section is difficult to estimate, but
is expected to be of the same order or larger than in the deep
inelastic scattering \cite{bal93b}.

To summarise, we have shown that Mueller's corrections are likely
to cut off the IR divergent integrals over the instanton size
in the process of gluon jet production with large transverse
momentum, indicating that their role is more important than
 usually believed.
In  general, one may thus conjecture that in {\em  any} hard
process there is a well-defined nonperturbative contribution due
to small instantons with the size of order $\rho\sim 1/(Q\alpha_s(Q)$,
where $Q$ is the corresponding hard scale.
A search for instanton-induced effects in large $q_\perp$ reactions
may be most fruitful because of larger rates and smaller
backgrounds.

\bigskip
V.B. thanks Prof. J.C. Collins for the invitation and the
hospitality extended to him by the theory group of the
Penn State University, where this work has been started.
The work by I.B. was supported by the US Department of Energy
under the grant DE-F902-90ER-40577.

\end{document}